# Smart Antenna Based Broadband communication in
# Intelligent Transportation system

Sourav Dhar, Debdattta Kandar, Tanushree Bose and Rabindranath Bera
Sikkim Manipal Institute of Technology, Majhitar, Rangpo, East Sikkim-737132

*Abstract:* **This paper presents a review for the development of Intelligent Transportation System (ITS) world wide and the use of Smart Antennas in ITS. This review work also discusses the usual problems in ITS and proposes the solution of such problems using smart antennas.**

I. INTRODUCTION

In United States, there were 6279000 motor vehicle accidents that accounted for 41611 deaths in 1991[1]. More health care dollars are consumed treating crash victims than any other cause of illness or injury. The United States Department of Transportation has declared that the reduction of vehicular fatalities is its top priority [2]. This is true for all other countries as well. Also demand for voice, data and multimedia services while moving in car, these increase the importance of Broadband wireless systems [3]. So, Intelligent Transportation Systems (ITS) are part of the national strategy for improving the security, safety, efficiency and comfort of every nation. To be specific, remote sensing and vehicular communications are given highest priority for ITS.

Information system and communications technology are integrated to the transport infrastructure and vehicles to improve safety and reduce vehicle wear, transportation times, and fuel consumption etc. ITS have been the umbrella under which significant efforts have been conducted in research, development, testing, deployment and integration of advanced technologies to improve the measures of effectiveness of national highway network. It vary in technologies applied, from basic management systems such as car navigation, traffic signal control systems, container management systems, variable message signs, automatic number plate recognition or speed cameras to monitoring applications, such as security CCTV systems and to more advanced applications that integrate live data and feedback from a number of other sources, such as parking guidance and information systems, weather information, bridge deicing systems etc.

Technologies like Wireless communications**,** Computational technologies, Sensing technologies, Floating car data/floating cellular data, Inductive loop detection, Video vehicle detection are already implemented in ITS.

ITS, encompass a broad range of wireless and wire-line communications-based information, control and electronics technologies. Short-range communications (less than 500 yards) can be accomplished using IEEE 802.11 protocols, specifically WAVE or the Dedicated Short Range Communications(DSRC) standard being promoted by the Intelligent Transportation Society of America and the United States Department of Transportation [4,5,6]. Theoretically, the range of these protocols can be extended using Mobile ad-hoc networks or Mesh networking [7]. Longer range communications have been proposed using infrastructure networks such as WiMAX (IEEE 802.16), Global System for Mobile Communications (GSM), or 3G [7]. Long-range communications using these methods are well established. The US FCC has allocated 75 MHz of spectrum in the 5.9 GHz band (5.8 GHz for Europe and Japan) [4] for DSRC to enhance the safety and productivity of the nation's transportation system.

Technological advances in telecommunications and information technology coupled with state-of-the-art microchip, RFID, and inexpensive intelligent beacon sensing technologies have enhanced the technical capabilities that will facilitate motorist safety benefits for Intelligent transportation systems globally. Sensing systems for ITS are vehicle and infrastructure based networked systems, e.g., intelligent vehicle technologies. The wide variety of remote sensors used in ITS applications (loops, probe vehicles, radar, cameras, etc.) is not as accurate as a stationary analyzer transportation system [8].

Broadband wireless systems play an increasingly important role in Intelligent Transportation Systems (ITS) by providing high speed wireless links between many ITS subsystems [9]. Smart antennas can greatly enhance the performance of wireless systems and fulfill the requirement of improving coverage range, capacity, data rate and quality of service [3]. Responsibility lies with the ITS designer to understand the working of a particular smart antenna before it is used for the intended operating environment. In the following sections we will discuss types and working of smart antennas and how they are used in Intelligent Transportation Systems.

II. FUNDAMENTALS OF SMART ANTENNAS

The term Smart Antenna is used in the wireless industry to represent many signal processing technologies that use multiple antennas on one or both ends of the wireless communication link. Smart antennas have superior capabilities to overcome the challenges for wireless communication systems. It can also provide array gain to increase range, diversity gain to improve performance under fading, and interference cancellation capabilities to increase capacity and to improve the quality of the wireless link [3]. Smart antennas can also be used to increase data rate, through delivery of higher SINR (Signal to Noise plus Interference Ratio) to the user or through spatial multiplexing [10]. It is necessary for systems designers to distinguish the different types of smart antennas and understand the benefits and limitations of each type. Depending on the number of transmitters and numbers of receivers in the wireless link, smart antenna system can be categorized as SIMO (Single Input Multiple Output), MISO (Multiple Input Single Output) and MIMO (Multiple Input Multiple Output).

Previously smart antennas are only implemented on the base stations side due to high cost and complexity in implementation. Hence, in satellite communication, SIMO techniques are used in uplink direction while MISO techniques are used in downlink direction [11]. For multiple receive, the receiver simply uses the signal from antenna with highest received signal power, which is termed as Signal Diversity [12]. Diversity gain is defined as the reduction of required SNR for a given BER (Bit Error Rate) under the given fading rate [13]. The maximum fading gain one can get by using multiple antennas is the single antenna fading margin, defined as the additional SNR required for a given BER under the given fading rate.

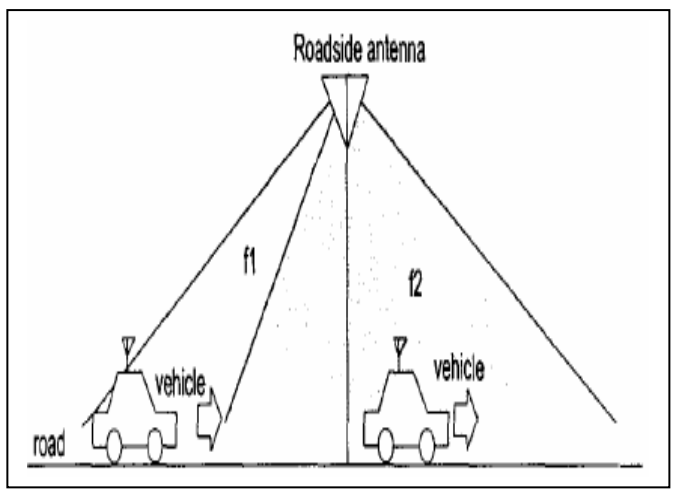

Fig. 1. Schematic drawing of a road to vehicle communication

Maximum Ratio Combining (MRC) smart antenna systems can get both array gain and diversity gain [3]. MRC combines the signals from the multiple antennas to maximize SNR (Signal to Noise ratio). Uplink array gain is defined as the increase in MRC receiver output SNR relative to that of a single antenna receiver [12]. The downlink combining gain is defined as the increase in power delivered to the user terminal receiver relative to that of a single antenna base station. MRC smart antenna systems have the tendency to lock on to strong interferers therefore they do not work well in situations where strong interferences exist.

Switched-beam systems can also provide diversity gain[14]. Switched-beam systems are simple to implement but has severe limitations. When angle spread is large, the diversity gain is limited because the system only selects signal from one beam instead of coherently combining the multiple-path components.

Fully Adaptive Array is another smart antenna which combines the signals from the multiple antennas to maximize SINR (Signal to Interference plus Noise ratio). Fully adaptive array not only gets the maximum diversity gain and array gain, but also cancels the interferences caused by different antenna elements. Due to its superior interference cancellation capability, fully adaptive array can reduce the frequency reuse (as low as 1) of cellular wireless systems [15], effectively increasing network capacity. With frequency reuse 1, frequency planning for wireless networks is greatly simplified. With fully adaptive array, spatial channels (two or more users sharing the same conventional channel), also known as Spatial Division Multiple Access can be implemented in the same cell, further increasing spectral efficiency.

III. CHALLENGES IN V2V COMMUNICATION

ROAD to vehicle communication system in ITS (Intelligent Transport Systems) are one of the important media, which can offer traffic safety and navigation information to drivers as well as entertainment information [7]. Fig.1 shows the schematic drawing of a road to vehicle communication scenario.

To offer Internet access service or a download service for large volume data files, long range communication is required. When constructing such a long-range communication zone (for example: several **km's** length) the following should be taken into account:

(i) Several services that have different communication systems or frequency bands will coexist in the same ITS communication network, and new services will be introduced one after another.

ii) The amount of communication traffic will change according to the continuous change in transportation traffic. Especially, communication traffic will drastically increase during traffic jams.

ii) Frequent handover will occur between adjacent. Spot zones due to the high speed of vehicles.

Regarding to (i) above, the Radio on Fiber (ROF) communication system [16] which conveys Radio Frequency (RF) signals through optical fibers, would be one of the key solutions to this. On the other hand, for (ii) and (iii) it is requested to construct an effective network that allows making efficient use of limited resources (frequency ha-lids) and creates smooth communication in a continuous **zone.** To meet **these requirements,** several technologies should be developed; that is, forecast of communication traffic according to transportation traffic, resource management technology, radio zone control technology and so on . In particular, to control a radio zone adaptively requires a roadside antenna to have the complex function of beam shaping and beam scanning. Therefore, it is important to develop such a smart antenna with a simple antenna configuration [17, 18].

IV. ADAPTIVE RADIO ZONE CONTROLS

*A. Zone Division and Zone Shift*

In this subsection, we will briefly explain two kinds of adaptive zone control schemes. One is a zone division scheme and the other is a zone shift scheme.

The zone division function divides one large radio zone into several small radio zones, depending on the vehicle traffic and the communication traffic. The system selects zone pattern #1 in Figure **2** that covers the whole radio zone adapted to the roadside antenna, in the case of small number of vehicles in one radio zone. On the other hand, zone pattern' *#2* or **#3** will be selected according to the volume of traffic under the antenna, during traffic jams. The zone division functions make it possible to assign several channels with different frequency channels adaptively; that is effective frequency reuse, as a result of which any driver can get the information channel from the roadside network independent of the traffic situation.

The zone shift function scans the radio zone (antenna beam) in accordance with the average speed of a vehicle group. This enables us to decrease the number of handovers within the specified continuous area. Fig. **3** explains how the zone shift works. Supposing, that the group including vehicle **A** runs from the left to the right in the Fig. 3. The radio zone with the frequency f l shifts its cover area at almost the same speed **as** that of vehicle **A** with the time interval T. Vehicle **A** will continue to communicate with only the channel f l and never experience or notice the handover within the zone shift area, so far as it runs at the same speed as the beam shift,. In the zone shift, there exists a case that, two adjacent antenna make **up** one radio zone (t = 1T to **3T in** Fig. **3).** In this case: the same signal should be transmitted from or received by the adjacent antennas along the road. Introducing the optical transmission/distribution system with ROF technology [19,20] could solve this.

*B. Beam Control Array Antenna for Adaptive Zone Control*

The development of a beam control array antenna is the primary issue to realize the adaptive radio zone control. The required functions for our beam control array antenna would be:
- To configure the radiation patterns suitable for the desired radio zone.
- To configure the radiation patterns with low side-lobe characteristics.
- To make switching time as small as possible in order to minimize the effect on the communication quality when the beam is switched.
- To have a simple interface with ROF access network.

Parameters for the road to vehicle communication system shown in Table I [21], are supposed for designing a beam control array antenna. The maximum radio zone length assigned to a roadside antenna is 100m, considering the expected radio zone for continuous road to vehicle communication systems. The maximum zone division number, and the input/output port number were determined as a minimum figure, in order to evaluate characteristics of the zone division and zone shift functions experimentally. As the maximum radio zone division number was determined to be four, the required beam patterns for a beam control array antenna to realize both zone division and zone shift demonstration would be ten patterns. Each beam pattern is depicted in Fig. 4.

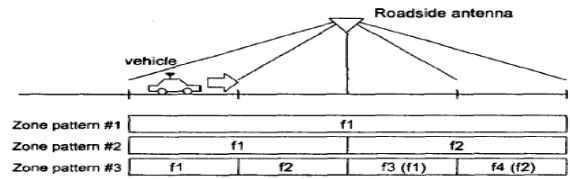

Fig. **2.** Schematic drawing of adaptive zone division

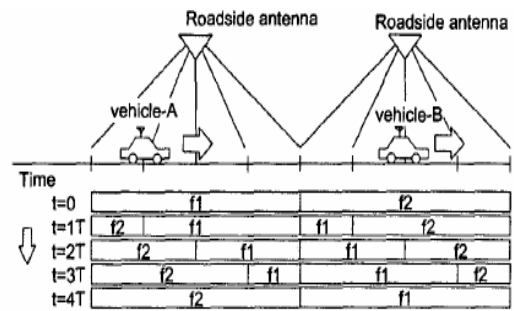

Fig. 3. Schematic drawing of adaptive zone shift.

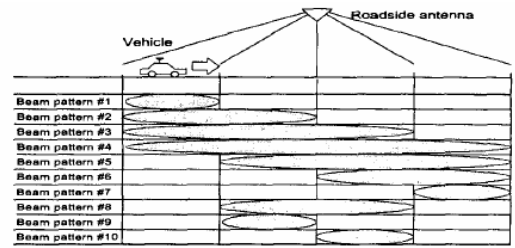

Fig. **4.** Required beam pattern for a beam control array antenna

TABLE I
PARAMETERS FOR MODEL COMMUNICATION SYSTEM

| Item | Values |
|---|---|
| Frequency band | 5.8 GHz band |
| Maximum zone length | 100 m |
| Maximum zone division | 4 |
| Minimum zone length | 25 m |
| I/O port number | 2 |
| Antenna height | 8 m (Roadside) |
|  | 1.5 m (Vehicle) |
| Required beam patterns | 10 patterns |

V. CONCLUSION

Smart antennas can greatly enhance the performance of wireless communication systems used in ITS. Smart antenna technology provides range extension, increased data rate, higher network capacity and better service quality. However, smart antenna represents many different ways of using multiple antennas on one or both ends of the wireless link. It is important to recognize the differences in performance among these smart antenna types. Moreover the paper shows that the antenna can change the radiation pattern, by adjusting only the weight of element beams used for road communication, which leads to simplifying and speeding up the beam control procedure. It can also integrate with ROF technologies for better road to road communication.

## REFERENCES

[1] Federal Communications Commission. Amendment of the commission's rules regarding dedicated short-range communication service in the 5.850-5.925 GHz band, FCC 02-302. Tech. rep., FCC, November 2002.
[2] J. Paniati (Dir., ITS, U.S Dept. of Transp.), Intelligent Safety Efforts in America, 10th ITS World Conf.
http://www.its.dot.gov/speeches/madridvii2003.ppt, Nov. 17, 2003.
[3] Xin Huang , "Smart Antennas for Intelligent Transportation Systems", 6th International Conference on ITS Telecommunications Proceedings, 2006.
[4]  USFCC, Report and Order, FCC 03-324, Dec. 2003.
[5] http://www.standards.its.dot.gov/Documents/dsrc_advisory.htm
[6]     Minutes   of   IEEE   DSRC   Standards   Group   meetings, http://www.leearmstrong.com/DSRC/DSRCHomeset.htm.
[7] http://en.wikipedia.org/wiki/Intelligent_transportation_system
[8] http://www.freepatentsonline.com/6804602.html
[9] "Broadband communication and its realization with broadband ISDN" Heinrich Armbruster and Gerhard Arndt, November 1087-Vol. 25, No. 11
IEEE Communications Magazine
[10] "Using Multistage Interference Cancellation Smart Antennas in Wideband CDMA Uplink", Hsin-Chin Liu and John F. Doherty, Signals, Systems and Computers, 2003. Conference Record of the Thirty-Seventh Asilomar Conference on Volume 1, Issue , 9-12 Nov. 2003 Page(s): 433 - 437 Vol.1
[11] http://www.nari.ee.ethz.ch/wireless/pubs/files/PIMRC2005.pdf
[12] M. Cooper, M. Goldberg, Intelligent Antennas: Spatial Division Multiple Access, IEC Annual Review of Communications, 1996,pp.999-1002.
[13] http://www.wirelessnetdesignline.com/showArticle.jhtml?articleID=161601515
[14] "Performance of Switched Beam Systems in Cellular Base Stations" Tushar Moorti and Arogyaswami Paulraj *Proceedings of ASILOMAR-29,* http://ieeexplore.ieee.org/iel3/3850/11240/00540577.pdf?tp=&isnumber=&arnumber=540577
[15] http://people.cornell.edu/pages/zf24/Adaptive_arrays.htm
[16]  For example, M. Fujise, H. Harada, K. Tokuda, and T. Ushikubo, "Development of PHS & ETC Dual-Service Radio on Fiber System at 5.8 GHz," in *Proc. 1999 Engineering Sciences Society Conference* of *IEICE,* A-17-3, Sept. 1999 .
[17] F. Dobias and W. Grabow, "Adaptive Array Antenna for 5.8 GHz Vehicle to Roadside Communication," in *Proc. 1994 IEEE 44th Vehicular Technology Conference,* vol. 3, pp. 1512-16, June 1994.
[18] F. Dobias and J. Gunther "Reconfigurable Array Antennas with Phase-only Control of Quantized Phase Shiften," in *Proc. 1995 IEEE 45th Vehicular Technology Conference,* vol. 1, pp. 35-39,1995.
[19] M. Yasunaga, Y. Okamoto, R. Miyamoto, and Y. Yamasaki, "Research Activities on Radio on Fiber Communication Network in TAO," in Proc. *ITST2000,* pp. 59-64, Oct. 2000.
[20] Y. Okamoto, R. Miyamoto, and M. Yasunaga, "ROF Access Transmission Systems for Road-Vehicle Communication." in *Proc.* of *the 2000 Engineering Sciences Society Conference of IEICE,* A-17-15, Sept. *2000*
[21] Y. Yamasaki, M. Yasunaga, Y. Murakami. and H. Moribe,  "A Beam Control Array Antenna for Road to Vehicle Communications", in 2001 IEEE Intelligent Transportation Systems Conference Proceedings - Oakland (CA) USA = August 25-29, 2001